\documentclass[a4paper,11pt]{article}
\usepackage{pos}
\usepackage{caption}
\usepackage{subcaption}
\usepackage{multirow}

\usepackage{etoolbox}  
\apptocmd{\thebibliography}{\setlength{\itemsep}{0pt}\setlength{\parskip}{0pt}}{}{}

\usepackage{comment}

\title{Joint eROSITA and H.E.S.S. analysis of MSH 15-52 using Gammapy}

\author*[a]{Katharina Egg}
\author[a]{Alison M. W. Mitchell}

\affiliation[a]{Erlangen Centre for Astroparticle Physics (ECAP), Friedrich-Alexander-Universität Erlangen-Nürnberg\\
   Nikolaus-Fiebiger-Straße 2, 91058 Erlangen}

\emailAdd{katharina.egg@fau.de}
\emailAdd{alison.mw.mitchell@fau.de}

\abstract{Pulsar wind nebulae (PWNe) are prominent sources in the very-high energy (VHE) gamma-ray sky, constituting the most numerous identified source class in the H.E.S.S. Galactic Plane Survey (HGPS). They are comprised of energetic particles originating from the pulsar and expanding into the surrounding medium. As such, PWNe are of very high scientific interest as PeVatron candidates, objects that could potentially accelerate particles up to PeV energies. Additionally other aspects of their acceleration mechanism are being actively investigated, such as the open question of whether they accelerate not only leptonic but also hadronic particles, and the details of their morphology and particle transport mechanism.
As PWNe emit photons over a broad range of the electromagnetic spectrum, multiwavelength (MWL) studies are crucial for the investigation and study of their emission.

In this vein we present a joint eROSITA X-ray and H.E.S.S. gamma-ray study of the PWN MSH 15-52. We showcase our custom code for integrating the EDR and DR1 eROSITA data into the Gammapy framework, a python package optimised for the analysis of gamma-ray data. We present the first 3D (spatial and spectral) fit to eROSITA data by using Gammapy.
We furthermore combine these data with the public H.E.S.S. gamma-ray observations of MSH 15-52, resulting in a joint physical fit of the underlying particle population, and a subsequent discussion of the physical implications of our results.
Finally we give an outlook towards future efforts in MWL studies of PWNe and the broader context of MWL data analysis with Gammapy.}

\ConferenceLogo{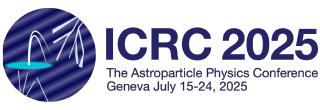}

\FullConference{39th International Cosmic Ray Conference (ICRC2025)\\
 15–24 July 2025\\
Geneva, Switzerland\\}

\begin{document}
\maketitle

\section{Introduction}

Pulsar wind nebulae (PWNe) 
form around pulsars; their strong magnetic fields and fast rotation accelerate particles to relativistic energies. An outflow of mainly electrons and positrons from the vicinity of the pulsar is created, interactions of these particles elicit the emission of photons, making PWNe observable throughout the electromagnetic spectrum \cite{Gaensler_2006,Mitchell_2022}.

PWNe are especially prominent sources at very-high energy (VHE), i.e. $E \geq 100 \, \mathrm{GeV}$, gamma-ray energies and make up the most numerous identified source class in the H.E.S.S. Galactic Plane Survey \cite{HGPS_2018}.
The study of their MWL spectral energy distributions (SEDs) is crucial in determining the properties of their incident particle spectra, to constrain particle transport mechanisms, and to test whether PWNe might be PeVatrons, objects that accelerate particles up to PeV energies \cite{Mitchell_2022}.


Data of different instruments are often combined through the use of separately extracted flux points. A joint analysis, however, in which the data is combined on the event level within one software package, has many advantages, such as a more nuanced treatment of statistics, systematics, and absorption \cite{Rosillo_2024}.

In this vein we have developed a pipeline for integrating 3D (i.e. containing spectral and spatial dimensions) eROSITA X-ray data into the Gammapy framework \cite{Donath_2023} to enable joint 3D analyses at the event level.
In this work we show a physical joint MWL fit of the particle injection spectrum of the PWN MSH 15-52 in Gammapy.

\section{Data used in this analysis}
\label{sec:data}
The analyses conducted in this work made use of data spanning from X-ray up to VHE gamma-ray energies in a range of $\sim 0.2 \, \mathrm{keV}$ to hundreds of TeV.

eROSITA X-ray data from the first data release (DR1) of the German eROSITA consortium were used \cite{Merloni_2024}. eROSITA is sensitive at soft X-ray energies of $\sim 0.2$ to $10\, \mathrm{keV}$ \cite{Predehl_2021}. It is made up of seven telescope modules (TMs) recording in parallel; because TM5 and TM7 are affected by an optical light leak they were excluded from the analysis \cite{Predehl_2021}. The eROSITA DR1 includes the eROSITA Science Analysis Software System (eSASS), the standard eROSITA software \cite{Brunner_2022}.

Additionally the H.E.S.S. data on MSH 15-52 from the H.E.S.S. public data release \cite{hess_public_data} were used. H.E.S.S. data spans an energy range of $\sim 0.1$ to $100\, \mathrm{TeV}$.
Finally the Fermi 4FGL catalog \cite{4fgl} flux points of MSH 15-52 were used, encompassing energies in a range of $\sim 0.1\,$GeV to $\sim 0.1 \,$TeV.

\section{eROSITA data in Gammapy}

X-ray and gamma-ray telescopes are both photon-counting experiments, in which the flux $\phi$ can be related to the number of counts $N$ through \cite{Donath_2023}:
\begin{equation}
    N (p,E) \mathrm{d}p \mathrm{d}E= E_\text{disp} \cdot [ PSF \cdot ( A_\text{eff} \cdot t_\text{obs} \cdot \phi ) ] + Bkg(p,E) \cdot t_\text{obs}
\end{equation}

To enable a comprehensive view of the source flux $\phi$, both the events data ($N$) and the corresponding Instrument Response Functions (IRFs) ($E_\text{disp}$, $A_\text{eff}$, $PSF$, and $Bkg$) need to be included in the analysis and thus need to be brought into a Gammapy-compatible format.

\begin{figure}[b]
    \centering
    \includegraphics[width=0.5\linewidth]{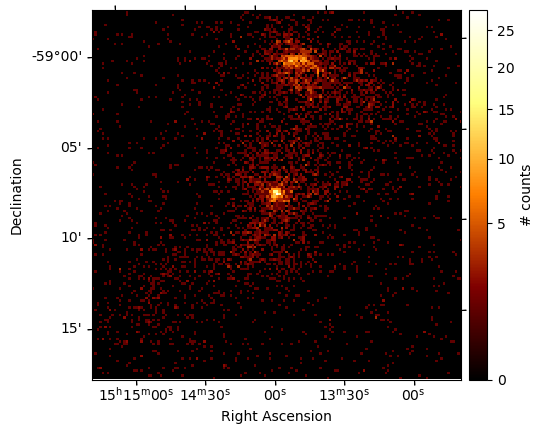}
    \caption{eROSITA DR1 stacked counts map of MSH 15-52}
    \label{fig:DR1_stacked}
\end{figure}

\subsection{Event data}

eROSITA event data is available in the form of eventfiles, available via the DR1\footnote{\url{https://erosita.mpe.mpg.de/dr1/}\label{foot:dr1}} or EDR\footnote{\url{https://erosita.mpe.mpg.de/edr/}} portal. A calibrated and filtered eventfile over a chosen circular region is created by running the eSASS command \texttt{evtool} \cite{Brunner_2022}. This eventfile contains an event list and a 2D image of the region as well as a GTI extension and a number of other extensions (containing e.g. eROSITA's pointing position over time).
Gammapy can parse this event list and its accompanying GTI extension once a few format changes have occurred, e.g. the \texttt{PI} column is renamed to \texttt{ENERGY}. A custom converter function was created to facilitate these changes.
A stacked eROSITA DR1 counts map of MSH 15-52 in Gammapy is shown in Figure \ref{fig:DR1_stacked}. This map contains data from all TMs not affected by the optical light leak\footnote{i.e. which eROSITA calls "820".}.

\subsection{Instrument Response Functions (IRFs) and Models}

The IRFs are similarly brought into a Gammapy-compatible format. Standard eROSITA spectra are accompanied by the Redistribution Matrix File (RMF) and the Ancillary Response File (ARF), containing the energy dispersion matrix and effective area respectively \cite{Brunner_2022}. For spectra of extended sources the Point Spread function (PSF) is usually neglected due to its small size (e.g. $16.1"$ half-energy width (HEW) at $1.5\, \mathrm{keV}$ \cite{Predehl_2021}). It is available on the DR1 portal$^{\ref{foot:dr1}}$ as a series of 2D images for different energy and offset bins.

\begin{figure}
    \centering
    \begin{subfigure}[b]{0.49\textwidth}
         \centering
         \includegraphics[width=\linewidth]{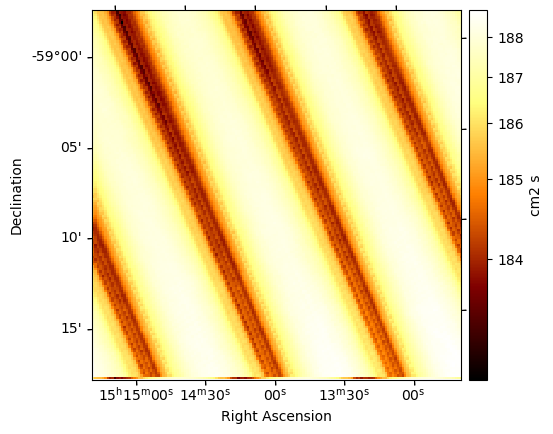}
    \end{subfigure}
     \hfill
     \begin{subfigure}[b]{0.49\textwidth}
    \includegraphics[width=\linewidth]{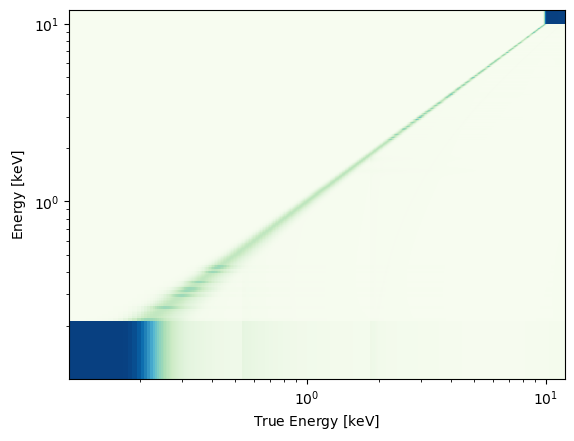}
    \end{subfigure}
    \caption{eROSITA IRFs in Gammapy, (left) exposure map, (right) energy dispersion matrix}
    \label{fig:irfs}  
\end{figure}

As the RMF is assumed to be constant for all eROSITA data with the same filtering, it is sufficient to read in a single file using the Gammapy \texttt{EDispKernel.read()} function. Due to a format issue another converter function was created. The energy dispersion matrix is visualized in the right panel of Figure \ref{fig:irfs}.

A 3D map of the effective area was created by repeatedly extracting the ARF over small regions (2x2 pixels or larger) with the eSASS command \texttt{srctool}. By multiplying the map with a map of the exposure time (created with the eSASS task \texttt{expmap}), a 3D exposure map is produced (see left panel of Figure \ref{fig:irfs}).

For the PSF two different scenarios have to be considered: for pointed observations (some EDR data) the radially averaged 2D PSF images can be directly used after converting them to Gammapy's \texttt{PSF3D} format. For scanning observations (DR1) a \texttt{PSFMap} is created instead by manually iterating over eROSITA's pointing positions over time and averaging the PSF profiles in each spatial, energy, and offset bin.

The background is treated with an OnOff approach: source regions are masked, then a 1D spectrum is extracted from the counts map and set as the background, while the acceptance compensates for region size and exposure time differences.
To import X-ray models into Gammapy via the sherpa package a modified version of the \texttt{SherpaWrapper} by \cite{Giunti_2022} was used.

\subsection{Validation}
\label{sec:validation}

3D eROSITA data in Gammapy was validated on MSH 15-52 by comparing standard eROSITA spectra to 1D spectra extracted from the 3D \texttt{MapDatasetOnOff} objects within Gammapy.
An absorbed powerlaw spectral model was fit to the Gammapy stacked spectrum using Gammapy as well as to the standard eROSITA spectrum (eSASS command \texttt{srctool}) using PyXspec \cite{PyXspec}. The spectra were extracted over a $10\, \mathrm{arcmin}$-radius around the central pulsar. The central pulsar, all point sources from the eROSITA DR1 one-band catalog \cite{Merloni_2024}, and a circular region around the close object RCW 89 were excluded.
The hydrogen absorption was modelled using the \texttt{TBabs} model \cite{Wilms_2000}, the nH parameter value was taken from \cite{HI4PI_2016}.
The results of this fit can be seen in Table \ref{tab:eROSITA_validation}, showing that the fit parameters agree within the $1 \sigma$ range.

%

\begin{table}
    \centering
    \begin{tabular}{l|cc}
    \hline
    \hline
    & \textbf{Gammapy} & \textbf{PyXspec}\\
    nH (frozen) & $1.360$ & $1.360$\\
    PhoIndex & $2.52 \pm  0.08$ & $2.58 \pm  0.08$\\
    norm & $0.022 \pm 0.001$ & $0.022 \pm 0.001$\\
    \hline
    \hline
    \end{tabular}
    \caption{Best fit parameters of absorbed powerlaw fit to 1D eROSITA spectrum extracted from a 3D dataset within Gammapy and to a standard eROSITA spectrum with PyXspec}
    \label{tab:eROSITA_validation}
\end{table}

\section{3D fit on eROSITA data}

Gammapy was used to conduct a 3D fit to the stacked eROSITA, combining a 2D spatial model and 1D spectral model. An absorbed powerlaw spectral model (as was used in Section \ref{sec:validation}) and a 2D Gaussian spatial model were chosen. Exclusion regions were applied as described in Section \ref{sec:validation} by setting a mask on the \texttt{MapDatasetOnOff}.

The fit converged successfully, showing that the shape of MSH 15-52 in the DR1 data can be approximated with a 2D Gaussian and for the first time simultaneously modelling both spectral and spatial parameters of eROSITA data. 
The best fit spatial model is shown on the counts map in Figure \ref{fig:MWL_regions} in gray, with parameters of this 3D fit given in Table \ref{tab:eROSITA_3D}.

One definite advantage of the 3D fit in Gammapy compared to a 1D fit is that, even with numerous exclusion regions, the full source flux is estimated, instead of the flux of a (possibly incomplete) spectrum extraction region. For deeper observations and/or more complex morphology, more complex source models are necessary to provide a good approximation of the source morphology.

\begin{table}
    \centering
    \begin{tabular}{cccccccc}
    \hline
    \hline
    nH (frozen) &     PhoIndex &     norm &     RA &     Dec &     sigma &     e &     phi  \\
    \hline
    $1.360$   & $2.009$ & $0.041$ & $228.448 $ & $-59.108$ & $5.523$ & $0.895$ & $-39.23$   \\
     &$\pm  0.08$ & $ \pm 0.004$ & $\pm 0.01 ^\circ$ & $\pm 0.01 ^\circ$ & $ \pm 0.42 \, \mathrm{arcmin}$ & $ \pm 0.02$ & $ \pm 1.91 ^\circ$\\
    \hline
    \hline
    \end{tabular}
    \caption{Best fit parameters of 3D fit to stacked DR1 eROSITA data of MSH 15-52}
    \label{tab:eROSITA_3D}
\end{table}

\section{Joint multiwavelength physical 3D fit}

In view of the successful 3D fit on the eROSITA DR1 data, a joint modeling of the underlying particle population of the PWN was conducted.
The particle population was modeled using models from the naima package \cite{naima}, imported into Gammapy with its available naima wrapper.

A leptonic scenario was assumed, which has been shown to describe the emission from PWNe and MSH 15-52 in particular in previous works (see e.g. \cite{2005_msh_hess}). Since the public H.E.S.S. runs are a subset of the H.E.S.S. data used in \cite{2005_msh_hess}, the same distance of $5.2 \, \mathrm{kpc}$ was assumed. The B field value of $17\, \mu \mathrm{G}$ found by \cite{2005_msh_hess} was used as a starting value.
The radiation fields for the inverse compton emission were calculated using the radiation field model by \cite{2017_Popescu}.

The total spectral model used in the fit was created by adding a synchrotron and inverse compton model with a single underlying particle population (i.e. linked parameters). Afterwards hydrogen absorption was applied to the summed models, using the \texttt{TBabs} model \cite{Wilms_2000}. The total model can be seen in Equation \ref{eq:tbabs_syn_ic}.

\begin{equation}
    \phi_\mathrm{total} (E)= \mathrm{TBabs} \cdot \left( \phi_\mathrm{syn}(E) + \phi_\mathrm{IC}(E)  \right)
\label{eq:tbabs_syn_ic}
\end{equation}

Separate 2D Gaussian spatial models were used for the eROSITA and H.E.S.S. data, owing to the fact that the size of PWNe very often differs in X-rays and gamma-rays \cite{Mitchell_2022}.

eROSITA, H.E.S.S., and Fermi data were used in the joint fit, as described in Section \ref{sec:data}. The three separate datasets were combined in a \texttt{Datasets} object in Gammapy and jointly fit with the physical spectral model and spatial models.

\noindent Several different particle spectra were tested, with 
an ExponentialCutoffPowerlaw model preferred: 
\begin{equation}
    N(E) = A \cdot \left( \frac{E}{E_0} \right)^{- \alpha} \cdot \exp{ \left( - \frac{E}{E_\mathrm{cutoff}} \right) }
    \label{eq:ecpl}
\end{equation}

\begin{figure}
    \centering
    \includegraphics[width=0.75\linewidth]{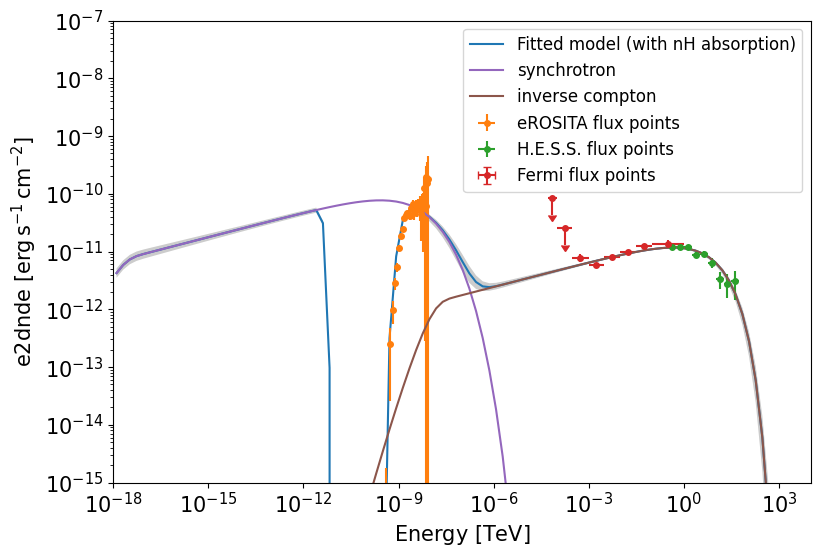}
    \caption{Physical fit to eROSITA 3D, H.E.S.S. 3D, and Fermi 1D flux point data with nH absorption (blue), synchrotron model (purple), inverse Compton model (brown), eROSITA flux points (orange), H.E.S.S. flux points (green), and Fermi 4FGL catalog flux points (red)}
    \label{fig:mwl_sed}
\end{figure}

\begin{figure}
    \centering
    \begin{subfigure}[b]{0.49\textwidth}
         \includegraphics[width=\linewidth]{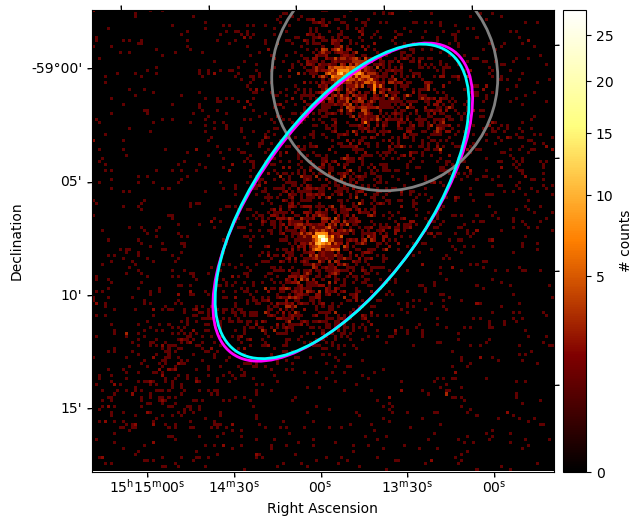}
    \end{subfigure}
     \hfill
     \begin{subfigure}[b]{0.49\textwidth}
    \includegraphics[width=\linewidth]{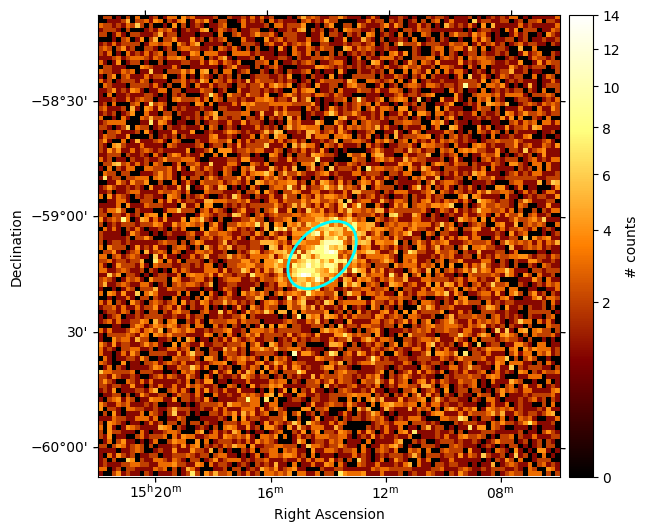}
    \end{subfigure}
    \caption{Regions resulting from the single instrument eROSITA fit (magenta), excluded region around RCW 89 (gray), and joint best fit regions shown in cyan over eROSITA (left) and H.E.S.S. counts map (right)}
    \label{fig:MWL_regions}
\end{figure}

Table \ref{tab:joint_fit} shows the best fit parameters, including a cut-off energy at $67\pm 13\,$TeV and magnetic field strength of $19.1 \pm 1.0 \mu \mathrm{G}$. 
The MWL spectral energy distribution (SED) 
in Figure \ref{fig:mwl_sed} 
includes flux points extracted from the eROSITA and H.E.S.S. datasets using the best fit model. The best fit regions superimposed over the counts maps can be seen in Figure \ref{fig:MWL_regions}.

\begin{table}
    \centering
    \begin{tabular}{lc|lcc}
    \hline
    \hline
   \multicolumn{2}{l|}{\textbf{Spectral:}}& \textbf{Spatial:} & \textbf{eROSITA} & \textbf{H.E.S.S.}\\
    nH (frozen) &  $1.360$&RA & $228.449 \pm 0.01 ^\circ$ & $228.55 \pm 0.01 ^\circ$\\
    Amplitude & $2.03 \pm 0.07 (10^{35} \, \mathrm{1/eV)}$& Dec & $-59.108 \pm 0.01 ^\circ$ & $-59.181 \pm 0.01 ^\circ$\\
    E\_0 (frozen) & $1.0 \, \mathrm{TeV}$& sigma & $5.405 \pm 0.33 \, \mathrm{arcmin}$ & $7.0 \pm 0.4 \, \mathrm{arcmin}$\\
    Alpha & $ 2.72 \pm 0.03$&e & $0.887 \pm 0.02$ & $0.77 \pm 0.05$\\
    E\_cutoff & $67 \pm 13 \, \mathrm{TeV}$&phi & $141.532 \pm 1.74 ^\circ$ &  $134.1 \pm 5.8 ^\circ$\\
    B & $19.1 \pm 1.0 \mu \, \mathrm{G}$&&&\\
    \hline
    \hline
    \end{tabular}
    \caption{Best fit parameters of physical fit to stacked 3D DR1 eROSITA data, 3D H.E.S.S. data, and Fermi 4FGL flux points of MSH 15-52}
    \label{tab:joint_fit}
\end{table}

\section{Conclusion and Outlook}

In this contribution we have presented a joint MWL physical fit to eROSITA, Fermi, and H.E.S.S. data to characterize the underlying particle population of the PWN MSH 15-52. We find the data well-described by a leptonic scenario, an ExponentialCutoffPowerlaw injection spectrum with a cutoff at $67\pm 13\,$TeV. The X-ray data allows us to characterize the magnetic field strength of the PWN, which we find 
to be $19.1 \pm 1.0 \mu \mathrm{G}$, 
in good agreement with previous studies (e.g. \cite{2005_msh_hess}).

Additionally we illustrate the steps towards this result and show our pipeline to integrate eROSITA data at the event level into Gammapy. We furthermore show the first 3D fit on eROSITA data in Gammapy.
The pipeline for integrating 3D eROSITA data into Gammapy will be made publicly available in the future for general use.
The MWL analysis on MSH 15-52 is also being continued with more data. 

\acknowledgments
{\footnotesize K. Egg and A. M. W. Mitchell are supported by the Deutsche Forschungsgemeinschaft, DFG project number 452934793.
This contribution was co-funded by a program supporting faculty-specific gender equality targets at Friedrich-Alexander University Erlangen-Nürnberg (FAU).

This work is based on data from eROSITA, the soft X-ray instrument aboard SRG, a joint Russian-German science mission supported by the Russian Space Agency (Roskosmos), in the interests of the Russian Academy of Sciences represented by its Space Research Institute (IKI), and the Deutsches Zentrum für Luft- und Raumfahrt (DLR). The SRG spacecraft was built by Lavochkin Association (NPOL) and its subcontractors, and is operated by NPOL with support from the Max Planck Institute for Extraterrestrial Physics (MPE). The development and construction of the eROSITA X-ray instrument was led by MPE, with contributions from the Dr. Karl Remeis Observatory Bamberg \& ECAP (FAU Erlangen-Nuernberg), the University of Hamburg Observatory, the Leibniz Institute for Astrophysics Potsdam (AIP), and the Institute for Astronomy and Astrophysics of the University of Tübingen, with the support of DLR and the Max Planck Society. The Argelander Institute for Astronomy of the University of Bonn and the Ludwig Maximilians Universität Munich also participated in the science preparation for eROSITA.
The eROSITA data shown here were processed using the eSASS software system developed by the German
eROSITA consortium.}

\bibliographystyle{JHEP}
\bibliography{bibliography.bib}

%

\end{document}